\documentclass[a4paper]{jpconf}

\usepackage{graphicx}
\usepackage{citesort}
\usepackage{amsmath}
\usepackage{bm}
\usepackage{mathtools}

\begin{document}

\title{Light Dark Matter and Superfluid He-4 from EFT}

\vspace{-0.5em}

\author{Andrea~Caputo$^1$, Angelo~Esposito$^2$ and Antonio~D.~Polosa$^3$}
\address{$^1$Instituto de Fisica Corpuscular, Universidad de Valencia and CSIC, Edificio Institutos Investigacion, Catedratico Jose Beltran 2, Paterna, 46980 Spain}

\address{$^2$Theoretical Particle Physics Laboratory (LPTP), Institute of Physics, EPFL, 1015 Lausanne, Switzerland}

\address{$^3$Dipartimento di Fisica and INFN, Sapienza Universit\`a di Roma, P.le Aldo Moro 2, I-00185 Roma, Italy}

\ead{angelo.esposito@epfl.ch}

\vspace{-0.5em}

\begin{abstract}
We study the response of a He-4 detector to the interaction of sub-GeV dark matter using an effective field theory for the superfluid. We compute the lifetime of the phonon, which agrees with what known from standard techniques, hence providing an important check of the effective field theory. We then study the process of emission of two phonons, and show how its rate is much more suppressed than the phase space expectations; this is a consequence of the conservation of the current associated to the superfluid symmetries.

\vspace{0.5em}

\noindent Talk presented at the TAUP 2019 conference.
\end{abstract}

\vspace{-3em}

\section{The EFT for the superfluid phonon and the dark matter}
The question about the nature of dark matter has been at the center of several theoretical and experimental efforts for decades. Recently, some attention has been devoted to the possibility of a dark matter lighter than the GeV, as suggested by different models---see e.g.~\cite{Boehm:2003hm,Hooper:2008im,Feng:2008ya,Hochberg:2014dra,Kuflik:2015isi,Kaplan:2009ag,Falkowski:2011xh,Hall:2009bx,Chu:2011be,Green:2017ybv} and also~\cite{Knapen:2017xzo,Bondarenko:2019vrb}. 
Experiments looking for particles in this mass range must have energy thresholds of an eV or below, and different ideas have been proposed~\cite{Hochberg:2015pha,Hochberg:2016ajh,Schutz:2016tid,Knapen:2016cue,Knapen:2017ekk,Dolan:2017xbu,Hochberg:2017wce,Griffin:2018bjn,Essig:2019xkx,Essig:2019kfe,Trickle:2019ovy,Emken:2019tni,Hochberg:2019cyy,Geilhufe:2019ndy,Cox:2019cod,Coskuner:2019odd,Sanchez-Martinez:2019bac,Acanfora:2019con,Caputo:2019cyg,Bunting:2017net,Campbell-Deem:2019hdx,Griffin:2019mvc,Trickle:2019nya}.

Here we study the possibility of detecting sub-GeV dark matter using superfluid He-4, in particular via the emission of collective excitations following the interaction of the dark matter with the bulk of the detector~\cite{Guo:2013dt,Schutz:2016tid,Knapen:2016cue,Acanfora:2019con,Caputo:2019cyg}. To do that, we employ a relativistic effective field theory (EFT) for superfluids~\cite{Son:2002zn,Nicolis:2011cs,Nicolis:2015sra}, which describes the phonon self-interactions as well as its interactions with the dark matter, starting solely from symmetry arguments~\cite{Acanfora:2019con,Caputo:2019cyg}.

From the EFT viewpoint a superfluid is a system with a conserved particle number, whose generator $Q$ is at finite density and spontaneously broken. Additionally, it also breaks boosts and time translations, generated by $H$, but preserves a particular combination of $H$ and $Q$~\cite{Nicolis:2015sra}. Given this symmetry breaking pattern the action describing the dynamics of the phonon and of the dark matter is completely determined---see~\cite{Acanfora:2019con,Caputo:2019cyg} for details. For the case of a scalar dark matter charged under a dark $U_\text{d}(1)$ group and interacting with the Standard Model via a scalar mediator coupled to the He-4 \emph{number density}, the interactions of interest are found to be
\begin{align} \label{eq:Seff}
\begin{split}
S_\text{eff}&=\int d^4x\bigg[\frac{1}{2}\dot\pi^2-\frac{c_s^2}{2}(\bm \nabla \pi)^2 -|\partial \chi|-m_\chi^2|\chi|^2 +\lambda_3\sqrt{\frac{m_\text{He}}{\bar n}}c_s\dot\pi(\bm \nabla \pi)^2+\lambda_3^\prime\sqrt{\frac{m_\text{He}}{\bar n}}c_s\dot\pi^3 \\
&\quad-\bigg(g_1\sqrt{\frac{m_\text{He}}{\bar n}}c_s\dot\pi-\frac{g_1}{2}\frac{c_s^2}{\bar n}(\bm \nabla\pi)^2+\frac{g_2}{2}\frac{m_\text{He} c_s^2}{\bar n}\dot\pi^2\bigg)|\chi|^2\bigg] \,,
\end{split}
\end{align}
with $\pi(x)$ and $\chi(x)$ being respectively the phonon and dark matter fields, and $c_s$, $\bar n$ and $m_\text{He}$ the He-4 equilibrium sound speed, number density and mass. Beside the unknown parameters of the dark sector, the effective couplings are completely determined by the superfluid equation of state, e.g. by $c_s$ as a function of the pressure $P$~\cite{Abraham:1970aa}. They are given by
\begin{align} \label{eq:couplings}
\begin{split}
\lambda_3=-\frac{1}{2m_\text{He}}\,, \quad \lambda_3^\prime=\frac{1}{6m_\text{He}c_s^2}-\frac{\bar n}{3c_s}\frac{dc_s}{dP}&\,, \quad g_1=-G_\chi m_\chi \frac{\bar n}{m_\text{He}c_s^2}\,, \\ 
 \quad g_2=-G_\chi m_\chi \frac{\bar n}{m_\text{He}c_s^2}\bigg(\frac{1}{m_\text{He}c_s^2}&-\frac{2\bar n}{c_s}\frac{dc_s}{dP}\bigg)\,.
 \end{split}
\end{align}
Here $G_\chi$ is an effective coupling of the dark sector with dimensions $(\text{mass})^{-2}$. The EFT can only describe the phonon degrees of freedom, and it has a momentum cutoff $\Lambda\simeq 1$~keV. Recall that the dispersion relation for an on-shell phonon of energy and momentum $(\omega,\bm q)$ is $\omega=c_sq$.

As a check one can compute the decay rate for a phonon of energy $\omega$, starting from the cubic interactions of Eq.~\eqref{eq:Seff}. Using the Feynman rules derived in~\cite{Acanfora:2019con} one easily finds
\begin{align}
\Gamma(\pi\to\pi\pi)=\frac{1}{240\pi c_s^5 m_\text{He} \bar n}\left(1+m_\text{He} \bar nc_s\frac{dc_s}{dP}\right)^2\omega^5\,,
\end{align}
in perfect agreement with what known from traditional methods~\cite{Maris:1977zz}.\footnote{Although all the excitations involved are gapless, the two final phonons are produced at some small but finite angle because of the mild momentum dependence of the sound speed~\cite{Maris:1977zz}.}

\vspace{-0.2em}

\section{Emission of one and two phonons}
Two different ways have been proposed to detect collective excitations in the superfluid He-4. The first one is via calorimetric techniques, and it requires the total energy deposited in the system to be larger than 1 meV~\cite{Hertel:2018aal}. The second one instead employs the so-called ``quantum evaporation'' and requires for the energy of the single excitations to be higher than 0.62 meV~\cite{Maris:2017xvi}.

From the action~\eqref{eq:Seff} one can compute the rates of emission of one and two phonons~\cite{Acanfora:2019con}. Being the velocity of the dark matter much larger than the speed of sound, $v_\chi\gg c_s$, the first emission happens at a fixed Cherenkov angle, $\cos\theta=\frac{c_s}{v_\chi}+\frac{q}{2m_\chi v_\chi}$, possibly allowing for directionality. However, it can only be detected via quantum evaporation, and it produces observable phonons only for a dark matter heavier than roughly 1 MeV.

\begin{figure}[t]
\begin{minipage}{0.45\textwidth}
\includegraphics[width=\textwidth]{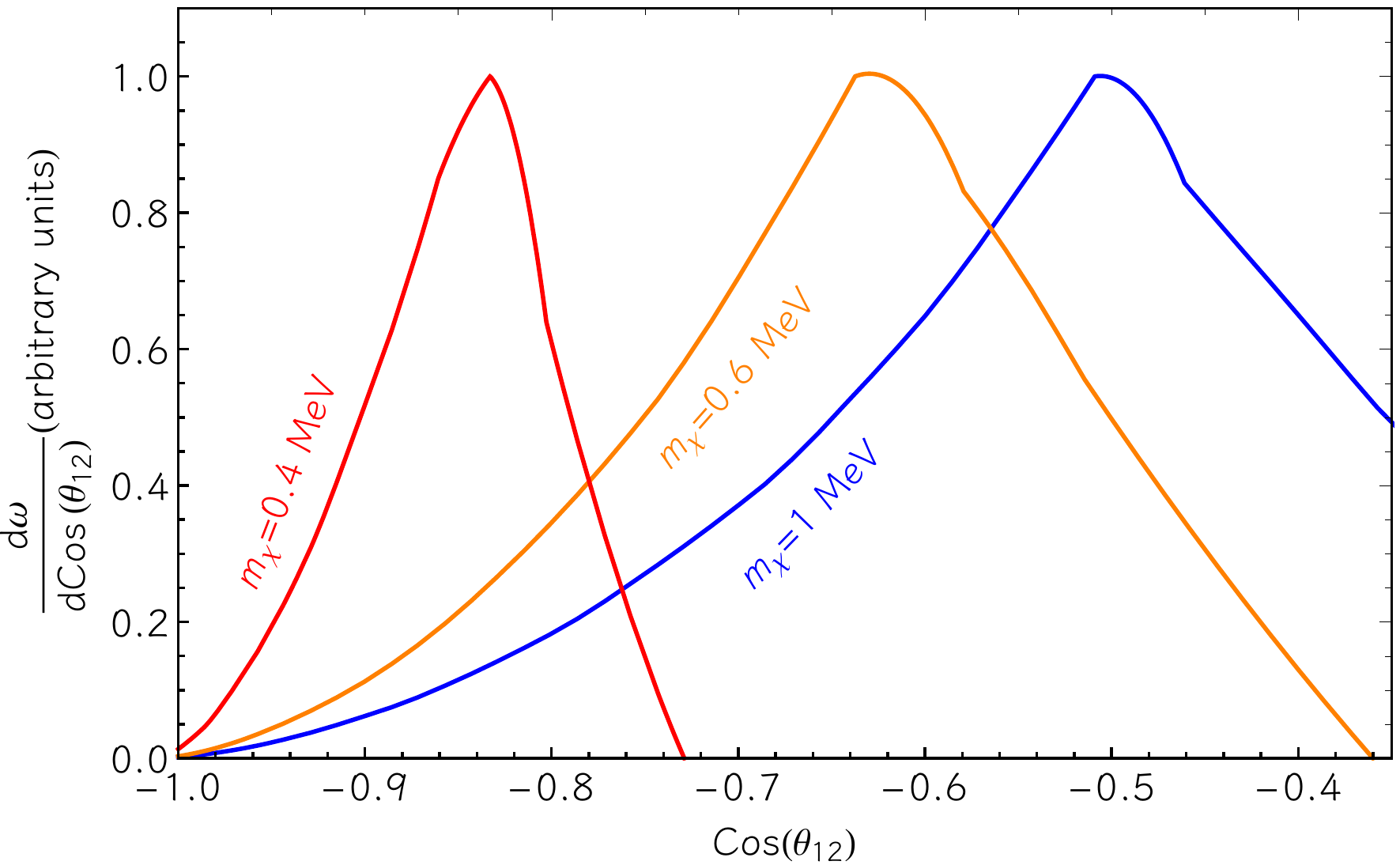}
\caption{\label{fig:dist} Angular distributions of the relative angle between the two final state phonons, $\theta_{12}$, assumed to be observed via quantum evaporation. When the dark matter become lighter the distributions are strongly peaked around $\theta_{12}\simeq \pi$. The curves have been normalized to have comparable amplitudes.}
\end{minipage}\hspace{2pc}%
\begin{minipage}{0.47\textwidth}
\includegraphics[width=\textwidth]{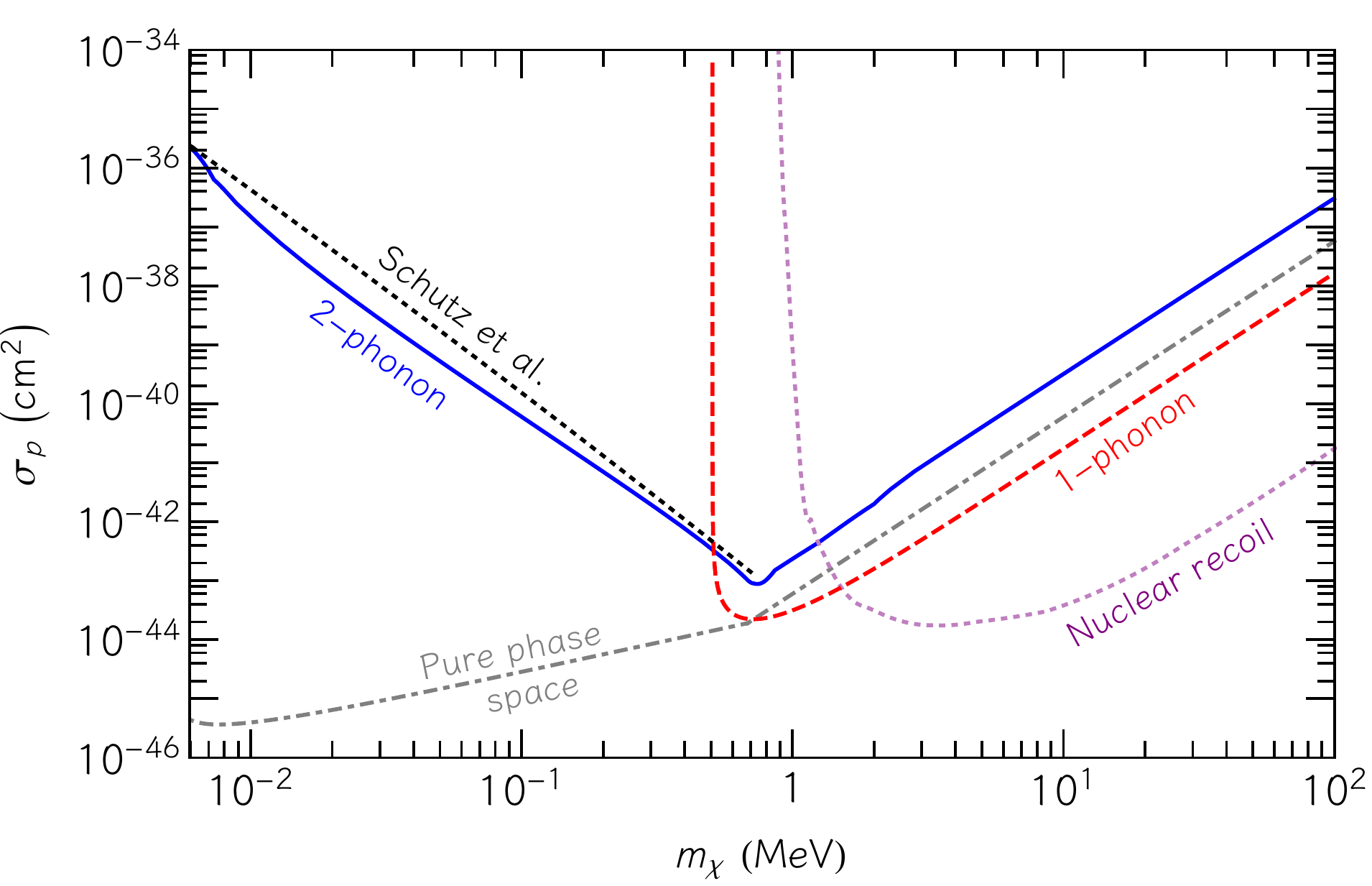}
\caption{\label{fig:sensitivity}Projected excluded region as referred to the 95\% C.L. for one kg of material and one year of exposure, assuming no background. For the emission of two phonons we assume detection via calorimetric techniques; for masses lighter than 1 MeV the curve obtained requiring quantum evaporation differs from the previous one by less than an order of magnitude.}
\end{minipage}  
\end{figure}

One the other hand, the process of emission of two phonons crucially allows to push the signal down to masses as light at the keV. In particular, for sub-MeV dark matter particles the dominant kinematical configuration is the one where the two outgoing phonons of momenta $\bm q_i$ are almost back-to-back, i.e. $\bm q_1\simeq-\bm q_2$ (see Figure~\ref{fig:dist}). In this configuration the energy released to the system is maximized and the final phonons could be detected with both calorimetric techniques as well as quantum evaporation. For details on the calculation of the rates described above we refer the interested reader to~\cite{Acanfora:2019con,Caputo:2019cyg}. From Figure~\ref{fig:dist} we see that the peak of the angular distribution is strongly dependent on the dark matter mass.

In Figure~\ref{fig:sensitivity} we instead report the projected sensitivity. Note that our results for the two-phonon emission are in agreement with those found in~\cite{Schutz:2016tid,Knapen:2016cue}, whose analysis includes both phonons and rotons. Given that our EFT describes only the former, we deduce that phonons contribute the most to the total rate of interest, with little role played by rotons.

\vspace{-0.2em}

\section{Rate suppression from conservation of current}

In Figure~\ref{fig:sensitivity} we have also reported a pure phase space exclusion plot for the two-phonon emission, obtained setting the matrix element to a constant, $\mathcal{M}=G_\chi m_\chi m_\text{He}c_s^2$, for dimensional reasons. As one can see, the two-phonon rate is suppressed by several orders of magnitude with respect to the naive phase space expectations. The reason for this was found in~\cite{Caputo:2019cyg}. In particular, as we have already explained, for a dark matter mass below the MeV, the dominant configuration is the one where the exchanged momentum, $\bm q=\bm q_1+\bm q_2$, is almost zero. It has been shown that in this regime there is a nontrivial cancellation between the two Feynman diagrams contributing to the process, such that the amplitude is identically zero in the strict $\bm q=0$ limit.\footnote{Remarkably, a similar cancellation happens for the emission of a single optical phonon by the dark matter in a crystal~\cite{Cox:2019cod}.} This however only happens if the dark matter is coupled to He-4 \emph{exactly} via the number density; if this coupling is modified even a bit the final rate increases by orders of magnitude~\cite{Caputo:2019cyg}.

Here we provide a further understanding of this effect, showing that it is nothing but a consequence of the conservation of current for the $U(1)$ symmetry of the superfluid. In fact, when the dark matter couples to the He-4 number density, the interaction Lagrangian is given by $\mathcal{L}_\text{int}=G_\chi m_\chi |\chi(x)|^2 n(x)$, where $n(x)$ is the local number density, i.e. including fluctuations (phonons)~\cite{Acanfora:2019con}. The initial and final states for the two-phonon emission process are $|i\rangle=|\chi(k)\rangle$ and $|f\rangle=|\chi(k^\prime)\pi(q_1)\pi(q_2)\rangle$. The contribution of the dark matter to the matrix element is trivial, and the only relevant part is the one involving the number density, $\langle \pi(q_1)\pi(q_2)|n(x)|\mu\rangle$, with $|\mu\rangle$ being the superfluid strongly interacting ground state. The number density is the time component of the conserved current associated with the superfluid $U(1)$, $J^0(x)=n(x)$. Consider then the LSZ formula for the matrix element of the current with two outgoing phonons:
\begin{align} \label{eq:LSZ}
\begin{split}
\langle\pi(q_1)\pi(q_2)|J^\mu(x)|\mu\rangle&=i^2\int d^4x_1d^4x_2 e^{-iq_1\cdot x_1-iq_2\cdot x_2}\big(\partial_{t_1}^2-c_s^2\bm\nabla_{x_1}^2\big)\big(\partial_{t_2}^2-c_s^2\bm\nabla_{x_2}^2\big) \times \\
&\qquad\times\langle\mu|T\big(\pi(x_1)\pi(x_2)J^\mu(x)\big)|\mu\rangle\,.
\end{split}
\end{align}
The phonon field shifts under the $U(1)$, $\pi\to\pi+a$, and the Ward identity for the conservation of current reads $\partial_\mu^{(x)}\langle\mu|T\big(\pi(x_1)\pi(x_2)J^\mu(x)\big)|\mu\rangle=-i\delta^4(x-x_1)\langle\mu|\pi(x_2)|\mu\rangle-i\delta^4(x-x_2)\langle\mu|\pi(x_1)|\mu\rangle$.
As usual, contact terms do not contribute to the on-shell matrix element since they shift the position of the pole on the external legs~\cite{Srednicki:2007qs}, and starting from Eq.~\eqref{eq:LSZ} one obtains, in momentum space, $q_\mu\langle\pi(q_1)\pi(q_2)|J^\mu(q)|\mu\rangle=0$. In the $\bm q=0$ limit one indeed gets than the matrix element for the emission of two phonons vanishes, $\langle \pi(\omega_1,\bm q_1)\pi(\omega_1,-\bm q_1)|n(\omega,\bm q=0)|\mu\rangle=0$. 

Note also that the argument above is easily generalized to more phonon fields. We then expect for this cancellation to happen for \emph{any} number of final state phonons in the limit of zero exchanged momentum.

\vspace{-0.2em}

\section{Conclusion}
Superfluid He-4 is a promising material to search for dark matter particles as light as the keV, with different ongoing R\&D efforts~\cite{Hertel:2018aal,Maris:2017xvi}. The EFTs for different states of matter can play a central role in the future theoretical understanding of such a system, as well as others.

\vspace{-0.2em}

\section*{References}
\bibliographystyle{iopart-num}
\bibliography{iopart-num}

\providecommand{\newblock}{}
\begin{thebibliography}{10}
\expandafter\ifx\csname url\endcsname\relax
  \def\url#1{{\tt #1}}\fi
\expandafter\ifx\csname urlprefix\endcsname\relax\def\urlprefix{URL }\fi
\providecommand{\eprint}[2][]{\url{#2}}

\bibitem{Boehm:2003hm}
Boehm C and Fayet P 2004 {\em Nucl. Phys.\/} {\bf B683} 219--263
  (\textit{Preprint} \eprint{hep-ph/0305261})

\bibitem{Hooper:2008im}
Hooper D and Zurek K~M 2008 {\em Phys. Rev.\/} {\bf D77} 087302
  (\textit{Preprint} \eprint{0801.3686})

\bibitem{Feng:2008ya}
Feng J~L and Kumar J 2008 {\em Phys. Rev. Lett.\/} {\bf 101} 231301
  (\textit{Preprint} \eprint{0803.4196})

\bibitem{Hochberg:2014dra}
Hochberg Y, Kuflik E, Volansky T and Wacker J~G 2014 {\em Phys. Rev. Lett.\/}
  {\bf 113} 171301 (\textit{Preprint} \eprint{1402.5143})

\bibitem{Kuflik:2015isi}
Kuflik E, Perelstein M, Lorier N~R~L and Tsai Y~D 2016 {\em Phys. Rev. Lett.\/}
  {\bf 116} 221302 (\textit{Preprint} \eprint{1512.04545})

\bibitem{Kaplan:2009ag}
Kaplan D~E, Luty M~A and Zurek K~M 2009 {\em Phys. Rev.\/} {\bf D79} 115016
  (\textit{Preprint} \eprint{0901.4117})

\bibitem{Falkowski:2011xh}
Falkowski A, Ruderman J~T and Volansky T 2011 {\em JHEP\/} {\bf 05} 106
  (\textit{Preprint} \eprint{1101.4936})

\bibitem{Hall:2009bx}
Hall L~J, Jedamzik K, March-Russell J and West S~M 2010 {\em JHEP\/} {\bf 03}
  080 (\textit{Preprint} \eprint{0911.1120})

\bibitem{Chu:2011be}
Chu X, Hambye T and Tytgat M~H~G 2012 {\em JCAP\/} {\bf 1205} 034
  (\textit{Preprint} \eprint{1112.0493})

\bibitem{Green:2017ybv}
Green D and Rajendran S 2017 {\em JHEP\/} {\bf 10} 013 (\textit{Preprint}
  \eprint{1701.08750})

\bibitem{Knapen:2017xzo}
Knapen S, Lin T and Zurek K~M 2017 {\em Phys. Rev.\/} {\bf D96} 115021
  (\textit{Preprint} \eprint{1709.07882})

\bibitem{Bondarenko:2019vrb}
Bondarenko K, Boyarsky A, Bringmann T, Hufnagel M, Schmidt-Hoberg K and
  Sokolenko A 2019  (\textit{Preprint} \eprint{1909.08632})

\bibitem{Hochberg:2015pha}
Hochberg Y, Zhao Y and Zurek K~M 2016 {\em Phys. Rev. Lett.\/} {\bf 116} 011301
  (\textit{Preprint} \eprint{1504.07237})

\bibitem{Hochberg:2016ajh}
Hochberg Y, Lin T and Zurek K~M 2016 {\em Phys. Rev.\/} {\bf D94} 015019
  (\textit{Preprint} \eprint{1604.06800})

\bibitem{Schutz:2016tid}
Schutz K and Zurek K~M 2016 {\em Phys. Rev. Lett.\/} {\bf 117} 121302
  (\textit{Preprint} \eprint{1604.08206})

\bibitem{Knapen:2016cue}
Knapen S, Lin T and Zurek K~M 2017 {\em Phys. Rev.\/} {\bf D95} 056019
  (\textit{Preprint} \eprint{1611.06228})

\bibitem{Knapen:2017ekk}
Knapen S, Lin T, Pyle M and Zurek K~M 2018 {\em Phys. Lett.\/} {\bf B785}
  386--390 (\textit{Preprint} \eprint{1712.06598})

\bibitem{Dolan:2017xbu}
Dolan M~J, Kahlhoefer F and McCabe C 2018 {\em Phys. Rev. Lett.\/} {\bf 121}
  101801 (\textit{Preprint} \eprint{1711.09906})

\bibitem{Hochberg:2017wce}
Hochberg Y, Kahn Y, Lisanti M, Zurek K~M, Grushin A~G, Ilan R, Griffin S~M, Liu
  Z~F, Weber S~F and Neaton J~B 2018 {\em Phys. Rev.\/} {\bf D97} 015004
  (\textit{Preprint} \eprint{1708.08929})

\bibitem{Griffin:2018bjn}
Griffin S, Knapen S, Lin T and Zurek K~M 2018 {\em Phys. Rev.\/} {\bf D98}
  115034 (\textit{Preprint} \eprint{1807.10291})

\bibitem{Essig:2019xkx}
Essig R, Pradler J, Sholapurkar M and Yu T~T 2019  (\textit{Preprint}
  \eprint{1908.10881})

\bibitem{Essig:2019kfe}
Essig R, Prez-Ros J, Ramani H and Slone O 2019  (\textit{Preprint}
  \eprint{1907.07682})

\bibitem{Trickle:2019ovy}
Trickle T, Zhang Z and Zurek K~M 2019  (\textit{Preprint} \eprint{1905.13744})

\bibitem{Emken:2019tni}
Emken T, Essig R, Kouvaris C and Sholapurkar M 2019 {\em JCAP\/} {\bf 1909} 070
  (\textit{Preprint} \eprint{1905.06348})

\bibitem{Hochberg:2019cyy}
Hochberg Y, Charaev I, Nam S~W, Verma V, Colangelo M and Berggren K~K 2019
  (\textit{Preprint} \eprint{1903.05101})

\bibitem{Geilhufe:2019ndy}
Geilhufe R~M, Kahlhoefer F and Winkler M~W 2019  (\textit{Preprint}
  \eprint{1910.02091})

\bibitem{Cox:2019cod}
Cox P, Melia T and Rajendran S 2019  (\textit{Preprint} \eprint{1905.05575})

\bibitem{Coskuner:2019odd}
Coskuner A, Mitridate A, Olivares A and Zurek K~M 2019  (\textit{Preprint}
  \eprint{1909.09170})

\bibitem{Sanchez-Martinez:2019bac}
Snchez-Martnez M~n, Robredo I, Bidauzarraga A, Bergara A, de~Juan F, Grushin
  A~G and Vergniory M~G 2019  (\textit{Preprint} \eprint{1905.04805})

\bibitem{Acanfora:2019con}
Acanfora F, Esposito A and Polosa A~D 2019 {\em Eur. Phys. J.\/} {\bf C79} 549
  (\textit{Preprint} \eprint{1902.02361})

\bibitem{Caputo:2019cyg}
Caputo A, Esposito A and Polosa A~D 2019  (\textit{Preprint}
  \eprint{1907.10635})

\bibitem{Bunting:2017net}
Bunting P~C, Gratta G, Melia T and Rajendran S 2017 {\em Phys. Rev.\/} {\bf
  D95} 095001 (\textit{Preprint} \eprint{1701.06566})

\bibitem{Campbell-Deem:2019hdx}
Campbell-Deem B, Cox P, Knapen S, Lin T and Melia T 2019  (\textit{Preprint}
  \eprint{1911.03482})

\bibitem{Griffin:2019mvc}
Griffin S~M, Inzani K, Trickle T, Zhang Z and Zurek K~M 2019
  (\textit{Preprint} \eprint{1910.10716})

\bibitem{Trickle:2019nya}
Trickle T, Zhang Z, Zurek K~M, Inzani K and Griffin S 2019  (\textit{Preprint}
  \eprint{1910.08092})

\bibitem{Guo:2013dt}
Guo W and McKinsey D~N 2013 {\em Phys. Rev.\/} {\bf D87} 115001
  (\textit{Preprint} \eprint{1302.0534})

\bibitem{Son:2002zn}
Son D~T 2002  (\textit{Preprint} \eprint{hep-ph/0204199})

\bibitem{Nicolis:2011cs}
Nicolis A 2011  (\textit{Preprint} \eprint{1108.2513})

\bibitem{Nicolis:2015sra}
Nicolis A, Penco R, Piazza F and Rattazzi R 2015 {\em JHEP\/} {\bf 06} 155
  (\textit{Preprint} \eprint{1501.03845})

\bibitem{Abraham:1970aa}
Abraham B~M 1970 {\em Physical Review A\/} {\bf 1} 250--257

\bibitem{Maris:1977zz}
Maris H~J 1977 {\em Rev. Mod. Phys.\/} {\bf 49} 341--359

\bibitem{Hertel:2018aal}
Hertel S~A, Biekert A, Lin J, Velan V and McKinsey D~N 2018  (\textit{Preprint}
  \eprint{1810.06283})

\bibitem{Maris:2017xvi}
Maris H~J, Seidel G~M and Stein D 2017 {\em Phys. Rev. Lett.\/} {\bf 119}
  181303 (\textit{Preprint} \eprint{1706.00117})

\bibitem{Srednicki:2007qs}
Srednicki M 2007 {\em {Quantum field theory}\/} (Cambridge University Press)
  ISBN 9780521864497, 9780511267208

\end{thebibliography}

\end{document}